\documentclass[aps,pra,showpacs,preprint,superscriptaddress]{revtex4}
\usepackage{hyperref}
\usepackage{amsmath}
\usepackage{amssymb}
\usepackage{epsfig}
\usepackage{natbib}
\usepackage{bm}
\usepackage{graphicx,amssymb,dcolumn,bm,amsmath,subfigure}
\usepackage[sort&compress]{natbib}
\usepackage{amsmath}

\newcommand*{\be}
{\begin{equation}}
\newcommand*{\ee}
{\end{equation}}

\makeatletter
\def\fnum@figure#1{\figurename\nobreakspace\thefigure .
      \hspace{0.02em}}
\def\fnum@table#1{\tablename\nobreakspace\thetable
       \hspace{0.02em}}
\makeatother

\begin{document}

\preprint{Submitted to PRA}
\title{Symmetric and asymmetric solitons in a nonlocal nonlinear coupler}
\author{Xianling Shi,$^{1}$ Boris A. Malomed,$^{2}$ Fangwei Ye,$^{1\ast }${%
\thanks{%
fangweiye@sjtu.edu.cn}}and Xianfeng Chen }
\date{\today }

\begin{abstract}
We study effects of nonlocality of the cubic self-focusing
nonlinearity on the stability and symmetry-breaking bifurcation
(SBB) of solitons in the model of a planar dual-core optical
waveguide with nonlocal (thermal) nonlinearity. In comparison with
the well-known coupled systems with the local nonlinearity, the
present setting is affected by the competition of different spatial
scales, \textit{viz}., the coupling length and correlation radius of
the nonlocality, $\sqrt{d}$. By means of numerical methods and
variational approximation (VA, which is relevant for small $d$), we
find that, with the increase of the correlation radius, the SBB
changes from subcritical into supercritical, which makes all the
asymmetric solitons stable. On the other hand, the nonlocality has
little influence on the stability of antisymmetric solitons.
Analytical results for the SBB are also obtained (actually, for
antisymmetric ``accessible solitons") in the opposite limit of the
ultra-nonlocal nonlinearity, using a coupler based on the
Snyder-Mitchell model. The results help to grasp the general picture
of the symmetry breaking in nonlocal couplers.
\end{abstract}

\maketitle

\address{ Department of Physics, The State Key laboratory on Fiber
Optic Local Area Communication Networks and Advanced Optical
Communication Systems, Shanghai Jiao Tong University, Shanghai
200240, China \\
Department of Physical Electronics, School of Electrical
Engineering, Faculty of Engineering, Tel Aviv University, Tel Aviv
69978, Israel}

{\ \renewcommand{\thefootnote}{\fnsymbol{footnote}} \footnotetext[1]{%
fangweiye@sjtu.edu.cn} }

\address{ Department of Physics, The State Key laboratory on Fiber
Optic Local Area Communication Networks and Advanced Optical
Communication Systems, Shanghai Jiao Tong University, Shanghai
200240, China \\
Department of Physical Electronics, School of Electrical
Engineering, Faculty of Engineering, Tel Aviv University, Tel Aviv
69978, Israel}


\section{Introduction}

\label{sec:sec1}Dual-core systems, featuring intrinsic nonlinearity in
parallel cores coupled by linear tunneling of wave fields, find their
realizations in various physical settings. Well-known systems of this type
in optics are twin-core fibers \cite{twin}-\cite{Snyder} (see also an early
review \cite{Wabnitz}) and Bragg gratings \cite{Mak}, as well as double
planar waveguides with the second-harmonic-generating intrinsic nonlinearity
\cite{Mak-chi2}. Similar settings for matter waves are represented by
two-layer Bose-Einstein condensates \cite{Arik}-\cite{Santos}. A fundamental
physical effect in nonlinear symmetric dual-core systems is the \textit{%
symmetry-breaking bifurcation} (SBB), alias the phase transition, which
destabilizes symmetric modes and gives rise to asymmetric ones. In nonlinear
optics, the SBB was studied in detail for continuous-wave (spatially
uniform) states \cite{Snyder} and solitons \cite{dual-core,Mak} in twin-core
fibers \cite{exact,Laval}, \cite{Maim}-\cite{1996} and Bragg gratings \cite%
{Mak} with the Kerr (cubic) nonlinearity, as well as for solitons in
double-core waveguides with the quadratic \cite{Mak-chi2} and cubic-quintic
\cite{Lior} nonlinearity. The SBB\ was studied too for matter-wave solitons
in two-layer BEC settings \cite{Arik,Warsaw}.

The self-focusing cubic nonlinearity gives rise to the SBB of the \textit{%
subcritical} type (alias the phase transition of the first kind) for
solitons in the symmetric dual-core system. The bifurcation of this type is
characterized by originally unstable branches of emerging asymmetric modes,
which at first extend backward (in the direction of weaker nonlinearity),
and then turn forward, retrieving the stability at the turning points \cite%
{bifurcations}. In this case, the system demonstrates a bistability and
hysteresis in a limited interval, characteristic to phase transitions of the
first kind. If the dual-core system is equipped with a periodic potential
(lattice) acting in the direction transversal to the propagation coordinate,
the character of the SBB changes to \textit{supercritical} above a certain
threshold value of the lattice's strength \cite{Warsaw}. The supercritical
bifurcation (alias the phase transition of the second kind) gives rise to
stable branches of asymmetric modes going in the forward direction \cite%
{bifurcations}. The SBBs belong to this type\ too in the twin-core Bragg
grating, and in quadratically nonlinear waveguides \cite{Mak,Mak-chi2}.

In addition to numerical analysis of symmetric, antisymmetric and asymmetric
soliton modes in dual-core system with the intrinsic cubic nonlinearity \cite%
{dual-core,anti}, the bifurcation point was found in an exact analytical
form \cite{exact}, and the emerging asymmetric solitons were studied in
detail by means of the variational approximation (VA) \cite%
{Laval,Maim,Pak,1996}. The latter method is relevant for studies of solitons
in many models originating in nonlinear optics and related fields \cite{VA},
while the possibility to find the exact bifurcation point is a feature
specific to particular systems.

The nonlinear response in optical media may feature spatial nonlocality,
which means that the local change of the refractive index induced by the
light beam depends on the distribution of the light intensity in a vicinity
of a given point \cite{ScienceNonlocal,ReviewNonlocal}. The nonlocality
arises when the nonlinear optical response involves mechanisms such as heat
diffusion, as analyzed theoretically \cite{Krolik} and demonstrated
experimentally \cite{MotiPRL,Canberra}, molecular reorientation in liquid
crystals \cite{AssantoPRL1, AssantoPRL2}, atomic diffusion \cite%
{SuterPRA,KrolikPRL,YePRA}, etc. The fields of nanophotonics and
plasmonics also give rise to effective nonlocalities, due to
light-matter interactions occurring in these media on deeply
subwavelength scales \cite{nano,Romania}.

Nonlocal nonlinearities are known in other physical media, including
plasmas \cite{plasma} and self-gravitating photonic beams
\cite{Rivlin}. Long-range interactions play an important role in
dipolar Bose-Einstein condensates (BECs) too \cite{DD}, and nonlocal
gravity-like interactions can be induced in BEC by means of laser
illumination \cite{quasi-gravity}.

The nonlocality, which introduces a new spatial scale, namely, the
correlation radius (denoted below as $\sqrt{d}$), may drastically alter
nonlinear excitations in optical systems, due to the interplay of $\sqrt{d}$%
with other natural scales. In particular, the nonlocality changes the
character of interactions between solitons \cite{AssantoOL}, and it
suppresses the beam's collapse and transverse instabilities \cite%
{KonotopPRE,OBangPRE}. The nonlocality also accounts for the formation of
new types of soliton modes \cite{YeOLTwin,YeOLBending}. However, to the best
of knowledge, the influence of nonlocality on the performance of optical
couplers has not been reported yet. In particular, new effects may be
expected due to the competition of $\sqrt{d}$ with the coupling length,
i.e., the interplay of nonlocal and local interactions. This is the
objective of the present work.

We consider the formation of solitons in a planar dual-core waveguide, in
which the nonlocal nonlinearity of the thermal type acts in both cores,
while the coupling between them remains linear and local, as the heat
diffusion does \ not transfer energy across the gap separating the
waveguides. Similar to couplers with the local nonlinearity, the nonlocal
model gives rise to three types of solitons, \textit{viz}., symmetric,
antisymmetric and asymmetric ones. However, the nonlocality significantly
affects the symmetry-breaking phase transition (SBB) for solitons, as well
as stability of the emerging asymmetric solitons, which are basic properties
of nonlinear couplers: at a critical value of the $\sqrt{d}$, the SBB
changes its character from sub- to supercritical. Taking into regard the
potential that nonlinear couplers have for various application to photonics,
such as all-optical switching \cite{switch,Wabnitz}, the use of the
nonlocality for the control of the soliton dynamics in these systems may
help to expand the range of the applications. While our analysis is
performed in terms of the thermal nonlinearity in optical waveguides, the
results may plausibly apply to other dual-core physical systems which
feature the nonlocal nonlinearity.

The paper is organized as follows. The model is formulated in Section II,
and analytical results are reported in Section III. These results are
obtained by means of the VA for solitons in the case of weak nonlocality
(small $\sqrt{d}$), and, on the other hand, the SBB is also investigated (in
fact, for antisymmetric solitons) in the opposite limit of the
ultra-nonlocal nonlinearity, in terms of a coupled system for
``accessible solitons" [the Snyder-Mitchell (SM) model \cite%
{ScienceNonlocal}]. In particular, the exact bifurcation point is
found for the SM system. The results for the small correlation
radius explicitly demonstrate the shift of the SBB point to larger
values of the soliton's power, and the trend to the transition of
the subcritical bifurcation into the supercritical one, while the
findings reported for the ultra-nonlocal system help to apprehend
the general situation. Numerical results, which provide the full
description of solitons in the nonlocal dual-core system for
moderate values of the correlation radius, are presented in Section
IV. In the case of the weak nonlocality, these results verify the
analytical results produced by the VA. The paper is concluded by
Section V.

\section{The model}

\label{sec:sec2} The propagation of optical beams along axis $z$ in the
planar dual-core waveguide with the intrinsic self-focusing nonlinearity of
the thermal type \cite{Krolik,ReviewNonlocal} is described by the system of
linearly coupled nonlinear Schr\"{o}dinger (NLS) equations for complex field
amplitudes $u$, $v$ in the two cores, and respective local perturbations $m,n
$ of the refractive index:
\begin{subequations}
\label{uvmn}
\begin{gather}
iu_{z}+\frac{1}{2}u_{xx}+mu+v=0, \\
iv_{z}+\frac{1}{2}v_{xx}+nv+u=0, \\
m-dm_{xx}=|u|^{2}, \\
n-dn_{xx}=|v|^{2},
\end{gather}%
where $x$ is the transverse coordinate, the coupling constant [the
coefficient in front of terms $v$ and $u$ in Eqs. (\ref{uvmn}a) and (\ref%
{uvmn}b), respectively] is scaled to be $1$ (accordingly, the coupling
length is also $\sim 1$), and $d$ is the squared correlation radius of the
nonlocality. In fact, $d$ controls the competition between the length scales
determined by the nonlocal and local interactions in the system.

Stationary solutions to Eqs. (\ref{uvmn}) with propagation constant $b$ are
looked for as
\end{subequations}
\begin{subequations}
\label{stat}
\begin{gather}
u\left( z,x\right) =e^{ibz}U(x),v\left( z,x\right) =e^{ibz}V(x), \\
m=m(x),n=n(x),
\end{gather}%
with real functions $U(x)$ and $V(x)$ obeying the following equations:
\end{subequations}
\begin{subequations}
\label{UV&mn}
\begin{gather}
-bU+\frac{1}{2}U^{\prime \prime }+mU+V=0, \\
-bV+\frac{1}{2}V^{\prime \prime }+nV+U=0, \\
m-dm^{\prime \prime }=U^{2}, \\
n-dn^{\prime \prime }=V^{2}.
\end{gather}%
Equations (\ref{uvmn}) conserve the total power,
\end{subequations}
\begin{equation}
P=P_{u}+P_{v}\equiv \int_{-\infty }^{\infty }|u|^{2}dx+\int_{-\infty
}^{\infty }|v|^{2}dx.  \label{P}
\end{equation}%
Obviously, symmetric [$U(x)=V(x)$] and antisymmetric $[U(x)=-V(x)]$ modes
have $P_{1}=P_{2}$, while asymmetric ones can be characterized by parameter
\begin{equation}
\Theta =\frac{P_{2}-P_{1}}{P_{2}+P_{1}},  \label{Theta}
\end{equation}%
which takes values $-1<\Theta <+1$.

Parallel to Eqs. (\ref{uvmn}), it is relevant to consider the ultra-nonlocal
model, taken in the form of two linearly coupled SM equations \cite%
{ScienceNonlocal},
\begin{subequations}
\label{SM}
\begin{gather}
iu_{z}+\frac{1}{2}u_{xx}-\frac{1}{2}P_{u}x^{2}u+v=0, \\
iv_{z}+\frac{1}{2}v_{xx}-\frac{1}{2}P_{v}x^{2}v+u=0,
\end{gather}%
where $P_{1,2}$ are the powers defined as per Eq. (\ref{P}). Actually, Eqs. (%
\ref{SM}) correspond to the version of Eqs. (\ref{uvmn}c) and (\ref{uvmn}d)
with spatially averaged right-hand sides. To the best of our knowledge, the
SM coupler was not considered before, while the extreme nonlocality
postulated in the SM model per se finds realizations and applications in
diverse optical \cite{SM-optical} and optomechanical \cite{SM-optomech}
settings.

\section{Analytical results}

\subsection{The variational approximation for the weakly nonlocal system}

To apply the VA to the present system, we note that, in the case of weak
nonlocality ($d\ll 1$), Eqs. (\ref{UV&mn}c) and (\ref{UV&mn}d) yield, in the
first approximation, $m=U^{2}+d\left( U^{2}\right) ^{\prime \prime }$, and $%
n=V^{2}+d\left( V^{2}\right) ^{\prime \prime }$ \cite{Krol}. The
substitution of this approximation into Eqs. (\ref{UV&mn}a) and (\ref{UV&mn}%
b) leads to a system of two coupled equations with nonlinear-diffraction
terms:
\end{subequations}
\begin{subequations}
\label{UV}
\begin{gather}
-bU+\frac{1}{2}U^{\prime \prime }+U^{3}+dU\left( U^{2}\right) ^{\prime
\prime }+V=0, \\
-bV+\frac{1}{2}V^{\prime \prime }+V^{3}+dV\left( V^{2}\right) ^{\prime
\prime }+U=0,
\end{gather}%
which may be derived from the Lagrangian with density
\end{subequations}
\begin{eqnarray}
\mathcal{L} &=&\frac{1}{4}\left[ \left( U^{\prime }\right) ^{2}+\left(
V^{\prime }\right) ^{2}\right] +\frac{b}{2}\left( U^{2}+V^{2}\right) -\frac{1%
}{4}\left( U^{4}+V^{4}\right)   \notag \\
&&+d\left[ U^{2}\left( U^{\prime }\right) ^{2}+V^{2}\left( V^{\prime
}\right) ^{2}\right] -UV.  \label{L}
\end{eqnarray}

The ansatz for soliton solutions may be naturally chosen as%
\begin{equation}
\left\{ U(x),V(x)\right\} =\left\{ A,B\right\} \mathrm{sech}\left(
x/W\right) ,  \label{ans}
\end{equation}%
where $A$ and $B$ are amplitudes of the two components, and $W$ is their
common width. The substitution of the ansatz into density (\ref{L}) and
evaluation of the integrals yields the corresponding Lagrangian,%
\begin{gather}
L\equiv \int_{-\infty }^{+\infty }\mathcal{L}dx=\frac{A^{2}+B^{2}}{6W}%
+b\left( A^{2}+B^{2}\right) W  \notag \\
-\frac{1}{3}\left( A^{4}+B^{4}\right) W+\frac{4d\left( A^{4}+B^{4}\right) }{%
15W}-2ABW.  \label{Leff}
\end{gather}%
This Lagrangian can be more conveniently rewritten in terms of the total
power $P$, see Eq. (\ref{P}), and \textit{power imbalance} $Q=P_{1}-P_{2}$,%
\begin{equation}
2\left( A^{2}+B^{2}\right) W\equiv P,~2\left( A^{2}-B^{2}\right) W\equiv Q,
\label{PQ}
\end{equation}%
as follows:%
\begin{gather}
2L=\frac{P}{6W^{2}}+bP-\frac{P^{2}+Q^{2}}{12W} \\
+\frac{d}{15}\frac{P^{2}+Q^{2}}{W^{3}}-\sigma \sqrt{P^{2}-Q^{2}},
\end{gather}%
where $\sigma =1$ for symmetric solitons and asymmetric ones generated from
them by the SBB, and $\sigma =-1$ for antisymmetric solitons. The
corresponding Euler-Lagrange equations are $\partial L/\partial W=\partial
L/\partial Q=\partial L/\partial P=0$, i.e.,%
\begin{gather}
-\frac{P}{W}+\frac{P^{2}+Q^{2}}{4}-\frac{3d}{5}\frac{P^{2}+Q^{2}}{W^{2}}=0,
\label{W} \\
Q\left( -\frac{1}{6W}+\frac{2d}{15W^{3}}+\frac{\sigma }{\sqrt{P^{2}-Q^{2}}}%
\right) =0,  \label{Q} \\
b=\frac{P}{6W}-\frac{1}{6W^{2}}-\frac{2dP}{15W^{3}}+\frac{\sigma P}{\sqrt{%
P^{2}-Q^{2}}}  \label{k}
\end{gather}

Equation (\ref{k}), which determines the propagation constant, $b$, is
detached from Eqs. (\ref{W}) and (\ref{Q}). Equation (\ref{Q}) yields either
$Q=0$, which corresponds to symmetric and antisymmetric solitons, or
\begin{equation}
-\frac{1}{6W}+\frac{2d}{15W^{3}}+\frac{1}{\sqrt{P^{2}-Q^{2}}}=0
\label{asymm}
\end{equation}%
for asymmetric ones. Further, the expansion of Eqs. (\ref{W}) and (\ref{k})
for small $d$, i.e., the weak nonlocality, yields%
\begin{equation}
W\approx \frac{4}{P}+\frac{3d}{5}P,~b\approx \frac{1}{32}P^{2}+\sigma -\frac{%
d}{192}P^{4},  \label{small-d}
\end{equation}%
which predicts that, naturally, the nonlocality makes the soliton wider, for
given total power $P$. This is confirmed by the numerical solutions, as
shown below.

The most essential point is to find the critical power $P_{\mathrm{bif}}$,
at which the asymmetric solitons bifurcate from the symmetric ones. This
value is determined by a system of equations (\ref{W}) and (\ref{asymm}), in
which one should set $Q=0$. Further, using the assumption of the weak
nonlocality, i.e., small $d$, the ensuing solution for $P_{\mathrm{bif}}$\
can be expanded up to order $d$, which yields%
\begin{equation}
P_{\mathrm{bif}}=2\sqrt{6}+\left( 24\sqrt{6}/5\right) d.  \label{bif}
\end{equation}%
Note that, at $d=0$, Eq. (\ref{bif}) gives $P_{\mathrm{bif}}(d=0)=2\sqrt{6}$
\cite{1996}, which may be compared to the known exact result \cite{exact}, $%
\left( P_{\mathrm{bif}}\right) _{\mathrm{exact}}=8/\sqrt{3}$, the relative
error being $0.0\allowbreak 57$.

The VA predicts, as per Eq. (\ref{bif}), the \emph{increase} of the
soliton's power at the bifurcation point due to the weak nonlocality. To
compare the prediction with the numerical findings, we take the slope of the
$P_{\mathrm{bif}}(d)$ dependence at $d=0$, for which Eq. (\ref{bif}) yields
\begin{equation}
\left[ \frac{d(P_{\mathrm{bif}})}{d(d)}|_{d=0}\right] _{\mathrm{variational}%
}=24\sqrt{6}/5\approx \allowbreak 11.758.  \label{VA}
\end{equation}%
On the other hand, the same slope obtained from the numerical solution (see
the next section) is
\begin{equation}
\left[ \frac{d(P_{\mathrm{bif}})}{d(d)}|_{d=0}\right] _{\mathrm{numerical}%
}\approx 10.867,  \label{num}
\end{equation}%
the relative error of the VA prediction being $0.075$ (see Table 1).

It is also possible to find another critical power, $P_{\mathrm{th}}$, which
corresponds to the turning point (i.e., the stabilization threshold for
asymmetric solitons) on the dependence of the asymmetry parameter, $\Theta
\equiv Q/P$ [see Eq. (\ref{Theta})], on total power $P$. To this end, one
should obtain a dependence between $\Theta $ and $P$, eliminating $W$ from
Eqs. (\ref{W}) and (\ref{asymm}), and identifying $P_{\mathrm{th}}$ from
condition
\begin{equation}
\frac{dP}{d\Theta }=0.  \label{theta}
\end{equation}%
In the limit of $d=0$, the result produced by the VA is known \cite{1996}:%
\begin{equation}
\left( P_{\mathrm{th}}\right) _{d=0}=3\cdot 6^{1/4}\approx 4.\,\allowbreak
695,  \label{Pth}
\end{equation}%
the corresponding value of the asymmetry at the critical point being $\Theta
_{\mathrm{th}}=1/\sqrt{3}$. On the other hand, the numerically found
threshold power at $d=0$ is%
\begin{equation}
\left[ \left( P_{\mathrm{th}}\right) _{d=0}\right] _{\mathrm{num}}\simeq
4.548,  \label{numTh}
\end{equation}%
hence the relative error produced by the comparison of Eqs. (\ref{Pth}) and (%
\ref{numTh}) is $0.031$ (see Table 1).

Further, the expansion of Eqs. (\ref{W}), (\ref{asymm}) and (\ref{theta})
for small $d$ yields the following prediction for the slope of curve $P_{%
\mathrm{th}}(d)$ at $d=0$:%
\begin{equation}
\left[ \frac{d(P_{\mathrm{th}})}{d(d)}|_{d=0}\right] _{\mathrm{variational}}=%
\frac{16}{5}\cdot 6^{3/4}\approx 12.268,  \label{VA2}
\end{equation}%
while the numerically found counterpart of this value is
\begin{equation}
\left[ \frac{d(P_{\mathrm{th}})}{d(d)}|_{d=0}\right] _{\mathrm{num}}\simeq
13.827,  \label{num2}
\end{equation}%
hence the respective relative error is $0.127$ (see Table 1).
\begin{table}[tbp]
\caption{The comparison between the VA-predicted characteristics of the
symmetry-breaking bifurcation, in the local and weakly nonlocal systems, and
their numerically found counterparts.}%
\begin{tabular}[t]{|l|c|c|c|}
\hline
Parameter & VA & Numeric & $\frac{\text{VA-Numer}}{\text{VA}}$(\%) \\ \hline
$P_{\text{th}}|_{d=0}$ & $4.6953$ & $4.5484$ & $3.13$ \\ \hline
$P_{\text{bif}}|_{d=0}$ & $4.8989$ & $4.6188$ & $5.72$ \\ \hline
$\frac{d(P_{\mathrm{th}})}{d(d)}|_{d=0}$ & $12.2681$ & $13.8270$ & $-12.71$
\\ \hline
$\frac{d(P_{\mathrm{bif}})}{d(d)}|_{d=0}$ & $11.7575$ & $10.8666$ & $7.58$
\\ \hline
\end{tabular}%
\label{tab:tb1}
\end{table}

Finally, we note that the relation
\begin{equation}
\frac{d(P_{\mathrm{th}})}{d(d)}|_{d=0}>\frac{d(P_{\mathrm{bif}})}{d(d)}%
|_{d=0},  \label{>}
\end{equation}%
see Eqs. (\ref{VA2}) and (\ref{VA}), suggests that $P_{\mathrm{th}}$ and $P_{%
\mathrm{bif}}$ will eventually \emph{merge} into a \emph{single}
critical/threshold value, which implies the transition from the subcritical
bifurcation to the supercritical one, as confirmed by numerical results
displayed below.

\subsection{The coupler for ``accessible solitons" (the
Snyder-Mitchell model)}

In the opposite case of the ultra-nonlocal nonlinearity, substitution (\ref%
{stat}a) transforms coupled SM equations \cite{ScienceNonlocal} and
Eq. (\ref{P}) into their stationary versions:
\begin{subequations}
\label{SMstat}
\begin{gather}
-bU+\frac{1}{2}U^{\prime \prime }-\frac{1}{2}P_{u}x^{2}U+V=0, \\
-bV+\frac{1}{2}V^{\prime \prime }-\frac{1}{2}P_{v}x^{2}V+U=0,
\end{gather}%
\end{subequations}
\begin{equation}
P_{u}=\int_{-\infty }^{\infty }U^{2}(x)dx,~P_{v}=\int_{-\infty }^{\infty
}V^{2}(x)dx.  \label{PP}
\end{equation}%
In spite of the apparently simple form of Eqs. (\ref{SMstat}) and (\ref{PP}%
), it is not possible to find exact solutions for asymmetric solitons. A
solution can be obtained, by means of the WKB approximation, in the limit
case of the strong asymmetry, $P_{v}\ll P_{u}$. In this case, the $U$
component is tantamount to the ground-state wave function of the harmonic
oscillator (HO), with the corresponding HO length $L_{u}=P_{u}^{-1/4}$,
eigenvalue of the propagation constant $b=-\sqrt{P_{u}/2}$, and amplitude
\begin{equation}
U(x=0)=\pi ^{-1/4}P_{u}^{5/8},  \label{U}
\end{equation}
while the weak $V$ component develops a broad shape, with a small amplitude,
$V(x=0)\approx -\sqrt{2/\pi }P_{v}^{3/4}P_{u}^{-1/8}$, and large width, $%
L_{v}\approx 2\sqrt{\sqrt{P_{u}}/P_{v}}$. The wave function of the $V$%
-component can be written in a relatively simple explicit WKB form
in the ``resonant" case,
\begin{equation}
P_{v}=P_{u}/\left( 2\left( 2N+1\right) \right) ^{2},  \label{R}
\end{equation}
with large integer $N$, when the $(2N)$-th energy eigenvalue in the shallow
HO potential (assuming that $N=0$ corresponds to the ground state) in the $V$%
-component is matched to the ground-state eigenvalue of the HO in the $U$%
-component:%
\begin{gather}
V(x)=-\sqrt{\frac{2}{\pi }}\left( \frac{P_{v}^{3}}{\sqrt{P_{u}}-P_{v}x^{2}}%
\right) ^{1/4}  \notag \\
\times \cos \left\{ \frac{1}{2}\sqrt{P_{v}}\left[ \frac{\sqrt{P_{u}}}{P_{v}}%
\arcsin \left( \sqrt{\frac{P_{v}}{\sqrt{P_{u}}}}x\right) +x\sqrt{\frac{\sqrt{%
P_{u}}}{P_{v}}-x^{2}}\right] \right\} ,  \label{WKB}
\end{gather}%
at $x^{2}<\sqrt{P_{u}}/P_{v}$, and $V(x)=0$ at $x^{2}>\sqrt{P_{u}}/P_{v}$
[if resonance condition (\ref{R}) does not hold, the WKB expression (\ref%
{WKB}) needs a correction around the edge points, $x^{2}=\sqrt{P_{u}}/P_{v}$%
].

It follows from Eq. (\ref{SMstat}b) taken at the inflexion point ($V^{\prime
\prime }=0$) closest to $x=0$ that the strongly asymmetric mode has \emph{%
opposite signs} of $U(x=0)$ and $V(x=0)$ (as written in the above formulas),
i.e., this asymmetric state develops from the \emph{antisymmetric} one. The
respective point of the antisymmetry-breaking bifurcation can be found in an
exact form. To this end, a solution to Eqs. (\ref{SMstat}) near the
bifurcation point is looked for as%
\begin{equation}
\left\{ U(x),V(x)\right\} =\pm U_{0}\exp \left( -\frac{1}{2}\sqrt{\frac{P}{2}%
}x^{2}\right) +\delta U(x),  \label{delta}
\end{equation}%
where the propagation constant and amplitude of the lowest unperturbed
antisymmetric mode, with total power $P$ (in both components), are%
\begin{equation}
b=-1-(1/2)\sqrt{P/2},  \label{b}
\end{equation}
\begin{equation}
U_{0}=\pi ^{-1/4}\left( P/2\right) ^{5/8},  \label{U0}
\end{equation}%
cf. Eq. (\ref{U}), and an infinitesimal antisymmetry-breaking perturbation, $%
\delta U(x)$, obeys the equation following from the substitution of
expression (\ref{delta}) into Eqs. (\ref{SMstat}) and (\ref{PP}) and
subsequent linearization:
\begin{subequations}
\label{dU}
\begin{gather}
\left( 1-b\right) \delta U+\frac{1}{2}\delta U^{\prime \prime }-\frac{1}{4}%
Px^{2}\delta U=U_{0}\left( \delta P\right) x^{2}\exp \left( -\frac{1}{2}%
\sqrt{\frac{P}{2}}x^{2}\right) , \\
\delta P\equiv U_{0}\int_{-\infty }^{+\infty }\exp \left( -\frac{1}{2}\sqrt{%
\frac{P}{2}}x^{2}\right) \delta U(x)dx.
\end{gather}

A relevant solution to inhomogeneous equation (\ref{dU}a) can be found as
\end{subequations}
\begin{subequations}
\label{d}
\begin{eqnarray}
\delta U &=&\left( \delta _{0}+\frac{1}{2}\delta _{2}x^{2}\right) \exp
\left( -\frac{1}{2}\sqrt{\frac{P}{2}}x^{2}\right) , \\
\delta _{0} &=&\frac{U_{0}\delta P}{3\sqrt{P/2}-4},~\delta _{2}=-\frac{%
4U_{0}\delta P}{3\sqrt{P/2}-4}.
\end{eqnarray}%
Finally, substituting expressions (\ref{d}) into Eq. (\ref{dU}b) and
canceling $\delta P$ as a common factor, the self-consistency
condition yields a simple exact result for the total power at which
the increase of
the spontaneous breaking of the antisymmetry occurs: $P_{\mathrm{cr}}^{%
\mathrm{(antisymm)}}=8$.

\section{Numerical Results}

\label{sec:sec4} Numerical solution of Eqs. (\ref{UV&mn}) was performed by
means of the standard relaxation method. As predicted by the VA, three
soliton families, symmetric, asymmetric, and antisymmetric ones, persist in
the nonlocal system. The numerically found relation between the total power,
$P$, and propagation constant $b$ for symmetric and antisymmetric solutions
is shown in Figs. \ref{fig:fig1}. It is seen that $b$ monotonically grows
with $P$ at a fixed value of the nonlocality range, $\sqrt{d}$[which implies
that the solitons may be stable in terms of the Vakhitov-Kolokolov (VK)
criterion \cite{VK}], and $b$ decreases with $d$ at fixed $P$. Both these
properties are correctly predicted by the VA, see Eq. (\ref{small-d}). The
fact that all the curves originate, at $P=0$, from the same point, is
obvious, as it immediately follows from Eqs. (\ref{UV&mn}) that $%
\lim_{P\rightarrow 0}b(P)=\sigma \equiv \mathrm{sgn}\left( UV\right) $.

\begin{figure}[tbph]
\centering
\includegraphics[width=0.6\textwidth]{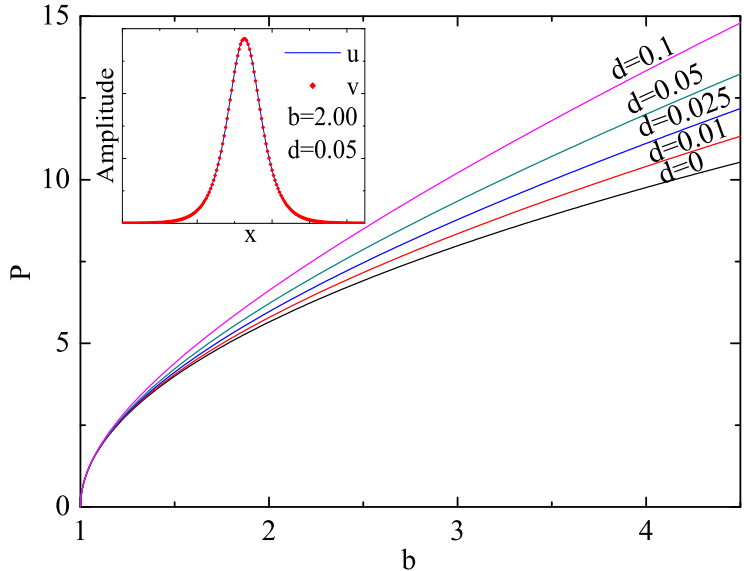}
\hspace{0\textwidth}
\caption{(Color online) Total power $P$ versus the soliton's propagation
constant, $b$, at different fixed values of the squared nonlocality
correlation radius, $d$, for symmetric solitons in the model based on Eqs. (%
\protect\ref{uvmn}). The inset shows a typical soliton profile. For
the antisymmetric solitons, $b$ is shifted by $\Delta b=-2$. All
quantities are plotted in arbitrary dimensionless units.}
\label{fig:fig1}
\end{figure}

Proceeding to numerically found asymmetric solitons, in Fig. \ref{fig:fig2}%
(a) we plot the respective $P(b)$ curves for for different fixed values of $d
$. As in the local system, asymmetric modes appear through the SBB when the
total power exceeds the threshold value, $P_{\mathrm{th}}$. Note that the
threshold, as well as the value of the total power at the bifurcation point,
$P=P_{\mathrm{bif}}$, significantly grow with $d$ [see Fig. \ref{fig:fig2}%
(b)], in accordance with the prediction of the VA given by Eqs. (\ref{Pth})
and (\ref{VA}). Further, the $P(b)$ curves change their shape with the
growth of the nonlocality radius: At small $d$, the slope, $dP/db$, is
initially negative (which definitely implies the instability, according to
the VK criterion \cite{VK}), going over to $dP/db>0$ with the further
increase of $b$. With the increase of $d$, the segment with the negative
slope shrinks, and disappears at $d>0.05$.

The change in the shape of the $P(b)$ characteristics is directly related to
the switch of the SBB from sub- to the supercritical type (in other words,
the switch from the symmetry-breaking phase transition from the first to
second kind) \cite{bifurcations}, as shown in Fig. \ref{fig:fig2}(c), where $%
P=P_{\mathrm{th}}$ determines the turning points of the $\Theta (P)$ curves,
and their unstable portions with $d\Theta /dP<0$ precisely correspond to the
segments with $dP/db<0$ in Fig. 2(b), both being confined to $P_{\mathrm{th}%
}<P<P_{\mathrm{bif}}$. Accordingly, the type of the SBB is subcritical, with
$P_{\mathrm{th}}<P_{\mathrm{bif}}$ at $d<0.05$, and supercritical, with $P_{%
\mathrm{th}}\equiv P_{\mathrm{bif}}$, at $d>0.05$. The merger of $P_{\mathrm{%
th}}$ and $P_{\mathrm{bif}}$ into the single value at $d>0.05$ is clearly
observed in Fig. \ref{fig:fig2}(b). Recall that, as mentioned above, the
trend to the merger of the two critical powers was predicted by the VA, see
Eq. (\ref{>}).

It is relevant to compare this result with the transition from the
subcritical SBB for solitons into the supercritical bifurcation under the
action of the periodic potential \cite{Warsaw}. Although the models are very
different (the one considered in Ref. \cite{Warsaw} is local), a common
feature is the introduction of a specific spatial scale---the nonlocality
range in the present model, $\sqrt{d}$, or the lattice period in the local
model---which is a factor accounting for the change of the character of the
SBB.

\begin{figure}[tbph]
\centering\subfigure{\label{fig:Fig.2a}\includegraphics[width=0.45%
\textwidth]{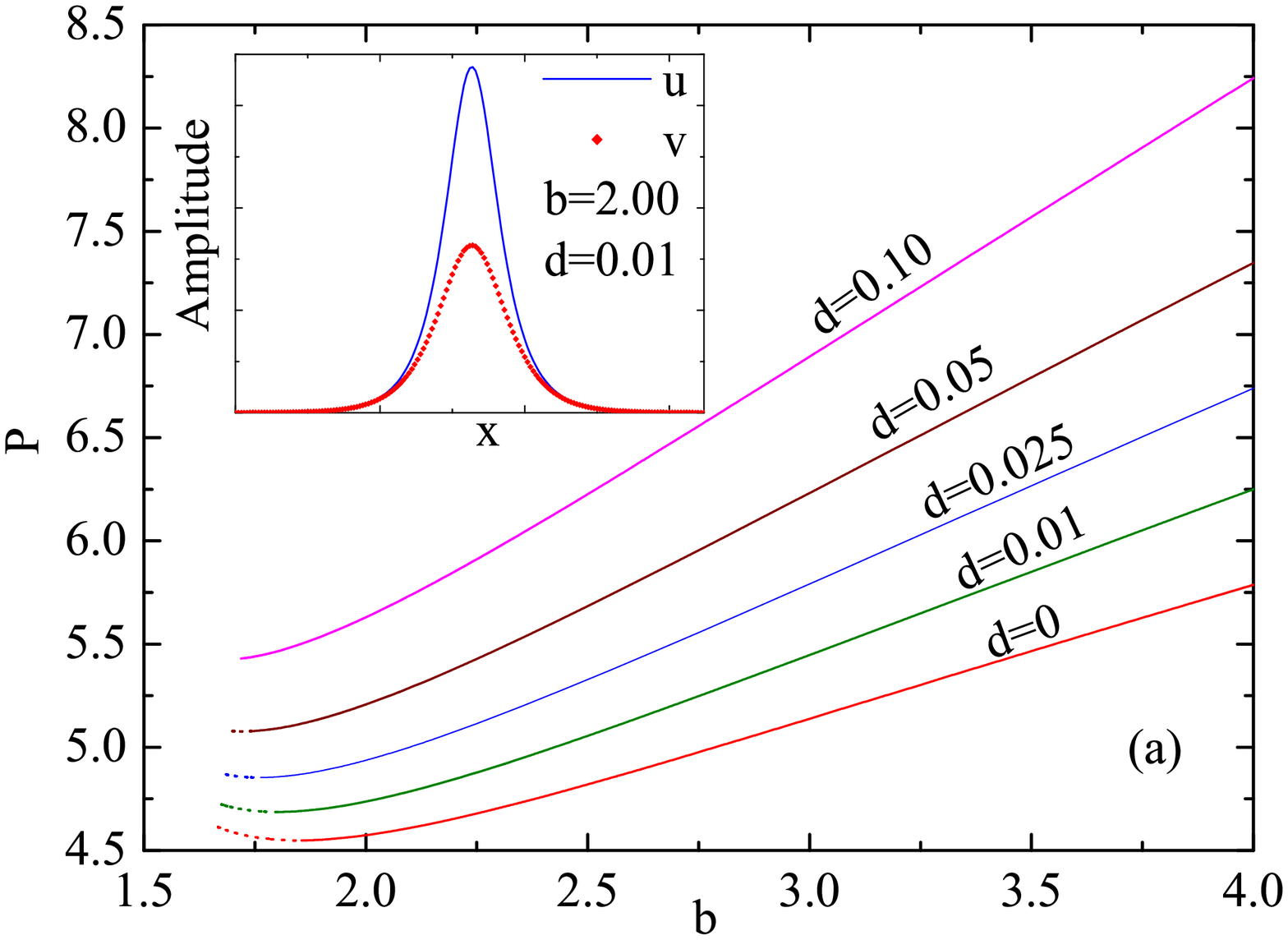}}\hspace{0pt}\subfigure{\label{fig:Fig.2b}%
\includegraphics[width=0.45\textwidth]{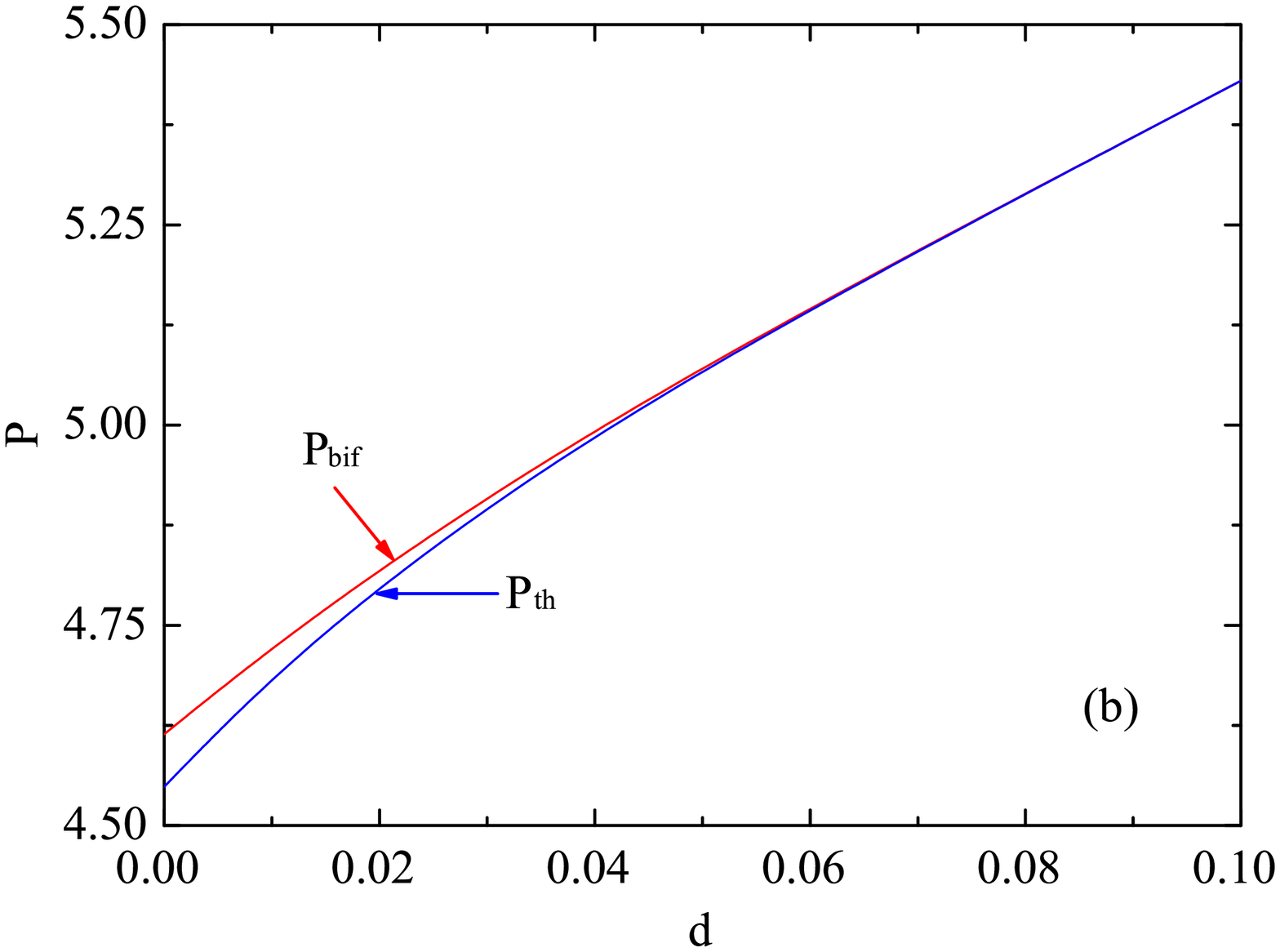}}\hspace{0pt}\newline
\subfigure{\label{fig:Fig.2b}\includegraphics[width=0.45%
\textwidth]{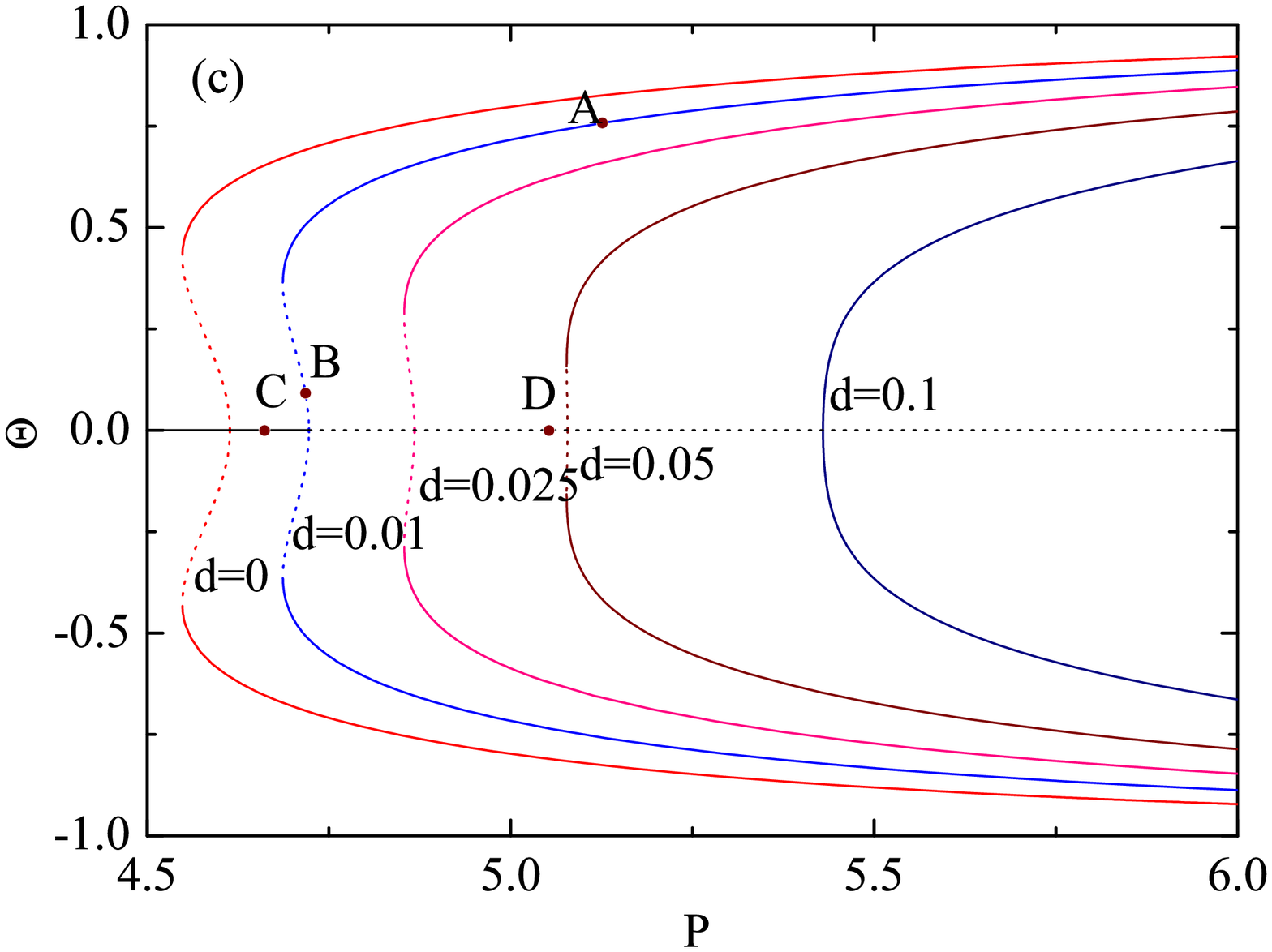}}\hspace{0pt} \caption{(Color online) (a)
Total power $P$ versus propagation constant $b$ for asymmetric
solitons at different values of the squared nonlocality radius, $d$.
The inset shows a typical soliton profile. (b) The dependence on
nonlocality $d$ of the total power, $P_{\mathrm{bif}}$, at which the
symmetry-breaking bifurcation gives rise to asymmetric solitons, and
of the threshold power, $P_{\mathrm{th}}$, at which the pair of
stable and unstable asymmetric solitons emerge subcritically. (c)
The bifurcation diagram accounting for the creation of the
asymmetric solitons from the symmetric ones. In panels (a) and (c),
dashed curves depict unstable portions of the asymmetric-soliton
families [the border between stable and unstable (dashed) parts of
the symmetric-soliton family in Fig. 2(c) corresponds to $d=0.01$].
All quantities are plotted in arbitrary dimensionless units.}
\label{fig:fig2}
\end{figure}




The stability of the solitons was tested by means of systematic simulations
of Eqs. (\ref{uvmn}), starting with perturbed initial conditions, $%
u(x,z=0)=U(x)(1+\rho (x)),v(x,z=0)=V(x)(1+\rho (x))$, where $U(x),V(x)$ is a
stationary solution, and $\rho (x)$ is a small-amplitude random function. As
expected, it has been found that the solid portions of the curves in Figs. %
\ref{fig:fig2}(a) and \ref{fig:fig2}(c), with $dP/db>0$ and $d\Theta /dP>0$,
carry stable solitons, while the dashed segments, with $dP/db<0$ and $%
d\Theta /dP<0$, represent unstable solutions. Thus, the increase of the
nonlocality radius, $\sqrt{d}$, gradually eliminates the instability region
for the asymmetric solitons, making them completely stable in the case when
the SBB is supercritical, i.e., at $d>0.05$.

It is relevant to explore the evolution of the two species of unstable
solitons in the dual-core system, \textit{viz}., asymmetric ones belonging
to the segments of the $\Theta (P)$ curves with the negative slope [i.e., $%
\Theta <\Theta (P_{\mathrm{th}})$, that exist at $d<0.05$], which are
represented, for example, by point B in Fig. \ref{fig:fig2}(c), and
symmetric solitons with $P>P_{\mathrm{bif}}$, sampled by point D in Fig. \ref%
{fig:fig2}(c). Figure \ref{fig:fig3}(a) displays the result for the unstable
asymmetric soliton, which demonstrates long-lived oscillations, initiated by
the instability, and eventual relaxation into a stable soliton with almost
the same power but higher asymmetry, $\Theta >\Theta (P_{\mathrm{th}})$,
which belongs to the stable branch of asymmetric modes in Fig. \ref{fig:fig2}%
(c). Further, Fig. \ref{fig:fig3}(b) demonstrates that the instability of
the symmetric soliton leads to its spontaneous rearrangement into an
asymmetric one, with nearly the same total power.

\begin{figure}[tbph]
\centering
\subfigure{\label{fig:Fig.3a}\includegraphics[width=0.7%
\textwidth]{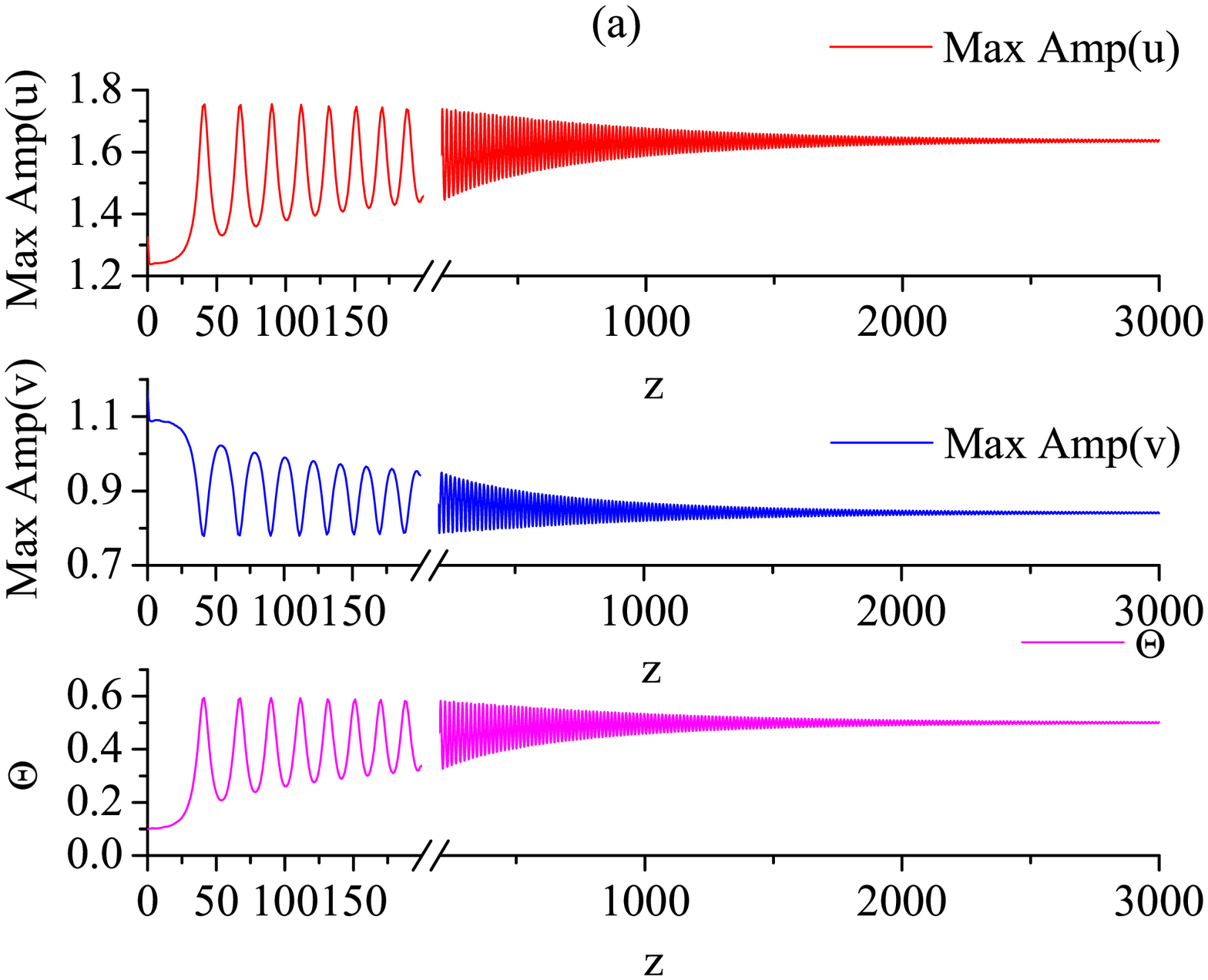}}\hspace{0\textwidth} \subfigure{\label{fig:Fig.3b}%
\includegraphics[width=0.7\textwidth]{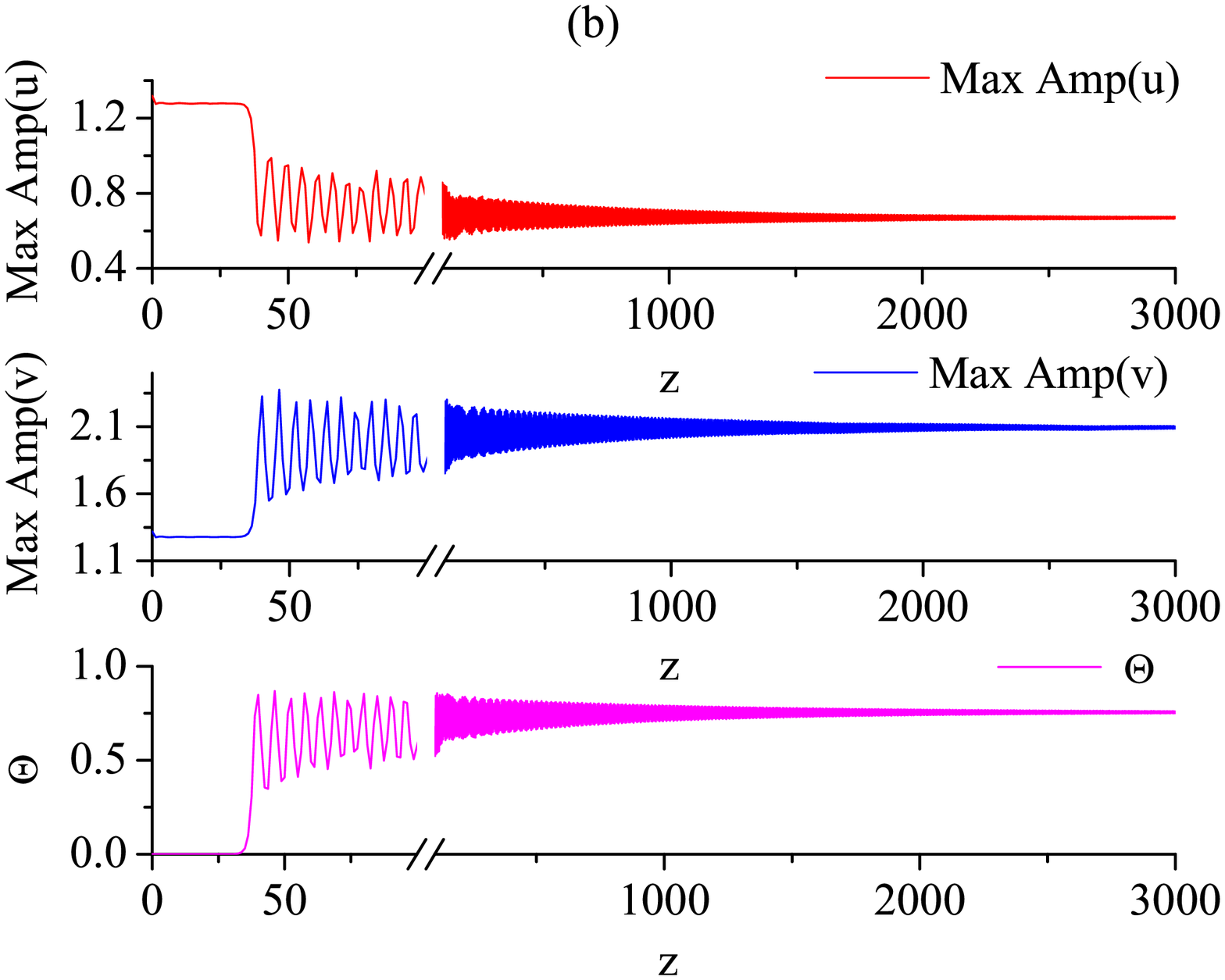}}\hspace{0\textwidth}
\caption{(Color online) (a,b): The evolution of perturbed unstable
solitons corresponding, respectively, to points B and D marked in
Fig. 2(c) (weakly asymmetric and symmetric solitons) is shown in
terms of amplitudes of both components, and asymmetry measure
(\protect\ref{Theta}). Both examples pertain to $d=0.01$. All
quantities are plotted in arbitrary dimensionless units.}
\label{fig:fig3}
\end{figure}

We have also studied the stability and evolution of antisymmetric solitons
for different strengths of the nonlocality in the model based on Eqs. (\ref%
{uvmn}) (the stability of the antisymmetric solitons in the model of the
coupler with the local nonlinearity was studied, in a numerical form, in
Ref. \cite{anti}). In contrast to the asymmetric solitons, where the
nonlocality leads to the transition from the subcritical SBB to the
supercritical bifurcation, and thus enhances the stability of the asymmetric
solitons, it has been found that the stability of the antisymmetric ones is
weakly affected by the nonlocality: the stability region slightly expands
under the action of the nonlocality, without dramatic changes.

\section{Conclusion}

\label{sec:sec5} We have introduced the nonlocal generalizations of
the standard model of the nonlinear directional coupler. The system
can be built, in particular, as a dual-core optical waveguide made
of a material with thermal nonlinearity. By means of the VA
(variational approximation) and systematic numerical analysis, we
have found that the relatively weak nonlocality shifts the SBB
(symmetry-breaking bifurcation) of solitons to larger values of the
total power, and eventually changes the character of the SBB from
subcritical to the supercritical (i.e., the corresponding phase
transition of the first kind goes over into the transition of the
second kind). Thus, the nonlocality of the cubic nonlinearity
enhances the stability for the asymmetric solitons, and eventually
leads to their stabilization in the whole existence domain, while
only slightly affecting the stability of antisymmetric solitons. For
the consideration of the opposite case of the ultra-nonlocal
nonlinearity, the coupler based on the SM (Snyder-Mitchell) model
was introduced. In that case, the phase transition leads to the
spontaneous breaking of the antisymmetry of the corresponding
two-component ``accessible solitons". The exact transition point was
found, and the strongly asymmetric states were found by means of the
WKB approximation.

The analysis reported in this paper can be extended in other directions. In
particular, as concerns nonlocal dual-core systems in other physical
contexts, it may be quite interesting to study the SBB and asymmetric
solitons in the case when the nonlocal interactions act between the cores,
an important example being a two-layer dipolar BEC \cite{Santos}. The
symmetry-breaking point can be easily found for the respectively modified SM
coupler model. A challenging extension is to construct two-dimensional
solitons in dual-core systems, where they may be stabilized against the
collapse by the nonlocality of the nonlinearity.


\section*{Acknowledgements}

F. Ye acknowledges the support of the National Natural Science Foundation of
China (Grant No. 10874119). B.A.M. appreciates hospitality of the Department
of Physics at the Shanghai Jiao Tong University, and of the Department of
Physics at East China Normal University (Shanghai).

\end{subequations}

\end{document}